\documentclass[article]{emulateapj}
\usepackage{natbib}
\usepackage{times}

\received{}
\accepted{}

\shorttitle{CO and N$_2$ desorption energies from water ice}
\shortauthors{Fayolle et al.}

\begin{document}

\title{CO and N$_2$ desorption energies from water ice. 

 }


\author{Edith C. Fayolle\altaffilmark{1},  Jodi Balfe\altaffilmark{1}, Ryan Loomis\altaffilmark{1}, Jennifer Bergner\altaffilmark{1}, Dawn Graninger\altaffilmark{1},  Mahesh Rajappan\altaffilmark{1}, and Karin I. \"Oberg\altaffilmark{1}}

 \altaffiltext{1}{Harvard-Smithsonian Center for Astrophysics, 60 Garden Street, Cambridge, MA 02138, USA}

\def\placetableone{

\begin{deluxetable*}{l c c c c c c c}[t]
\tablecaption{Coverages, desorption energy for the pure ice multilayer regime or mean desorption energy with full width half maximum for the sub-monolayer regime on water substrate, and the $^{15}$N$_2$ to $^{13}$CO desorption energy ratio for various substrates.}
\tablecolumns{8}
\tablewidth{0.8\textwidth}

\tablehead{
\colhead{Substrate} &  \multicolumn{2}{c}{$^{13}$CO}          &     \colhead{}             &  \multicolumn{2}{c}{$^{15}$N$_2$}& \colhead{}  & \colhead{E$_{\rm des}^{^{15}\rm N_2}$/E$_{\rm des}^{^{13}\rm CO}$} \\
\cline{2-3} \cline{5-6}\\
\colhead{}             &\colhead{Coverage / ML$_{eq}$}&\colhead{E$_{des}$ /K}& \colhead{}  & \colhead{Coverage / ML$_{eq}$}&\colhead{E$_{des}$ /K}& \colhead{} &\colhead{}}

\startdata

Pure ice \vspace{0.1 cm}&  5.0 & 866 $\pm$ 68\tablenotemark{*} & & 5.3 & 770 $\pm$ 68\tablenotemark{*} & & 0.89 $\pm$ 0.02 \\
H$_2$O (compact) \vspace{0.1 cm}&  1.3 & 1155 [133]& & 1.4 & 1034 [133] & & 0.90 $\pm$ 0.04\\
H$_2$O (compact) \vspace{0.1 cm}&  0.8 & 1180 [131]& & 0.7 & 1051 [127] & & 0.89 $\pm$ 0.04\\
H$_2$O (compact) \vspace{0.1 cm}&  0.3 & 1236 [139]& & 0.4 & 1090 [133] & & 0.88 $\pm$ 0.04\\
H$_2$O (compact) \vspace{0.1 cm}&  0.2 & 1298 [116]& & 0.2 & 1143 [113] & & 0.88 $\pm$ 0.04\\
H$_2$O (porous) &  0.7 & 1575 [117]& & 0.8 & 1435 [132] & & 0.91 $\pm$ 0.03

\enddata

\tablenotetext{*}{The pure ice desorption energies are obtained by a zeroth order fit and are given with uncertainties mainly due to the absolute error on the temperature}

\label{Tab_1}
\end{deluxetable*}

}


\def\placefigureone{

\begin{figure}[b]
  \centering
  \includegraphics[width=0.47\textwidth]{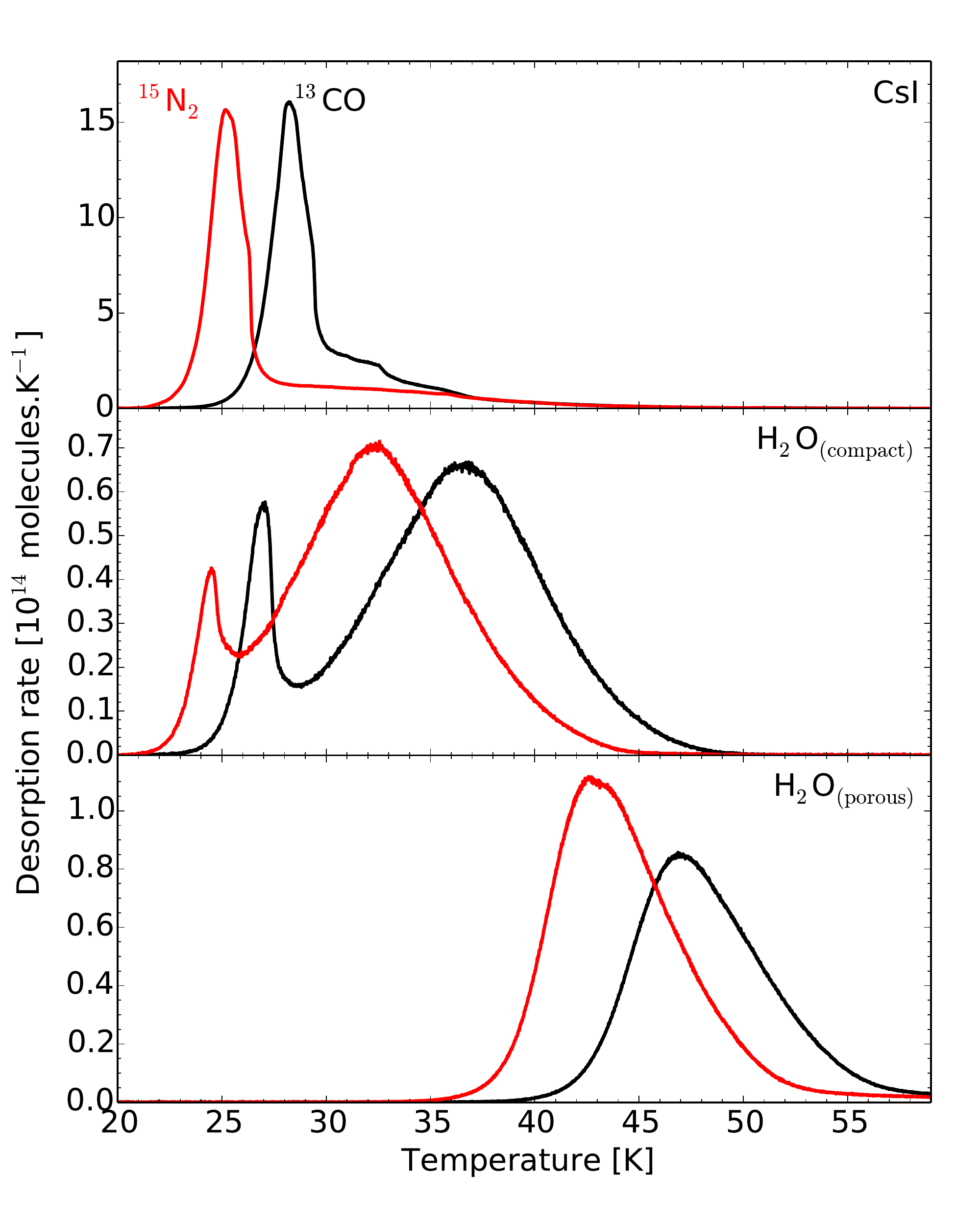}
  \caption{$^{13}$CO (solid black lines) and $^{15}$N$_2$ (solid red lines) TPD curves from pure ice and H$_2$O ice surface at 1 K min$^{-1}$. The upper panel presents the desorption of 5~ML of $^{13}$CO and 5~ML of $^{15}$N$_2$ deposited on a CsI window. The middle panel shows the desorption of 0.8~ML of $^{13}$CO and 0.7~ML of $^{15}$N$_2$ deposited on amorphous compact water (grown at 100~K). The lower panel shows the TPD curves of 2~ML $^{13}$CO and 2~ML of $^{15}$N$_2$ up to 65~K deposited on amorphous porous water (previously deposited at 14~K). For these later experiments, only 0.7~ML of $^{13}$CO and 0.8~ML of $^{15}$N$_2$ desorb below 65~K since a fraction of $^{13}$CO or $^{15}$N$_2$ stays trapped within the H$_2$O ice.}
  \label{Fig_1}
\end{figure} 

}

\def\placefiguretwo{
\begin{figure*}[t]
  \centering
  \includegraphics[width=0.9\textwidth]{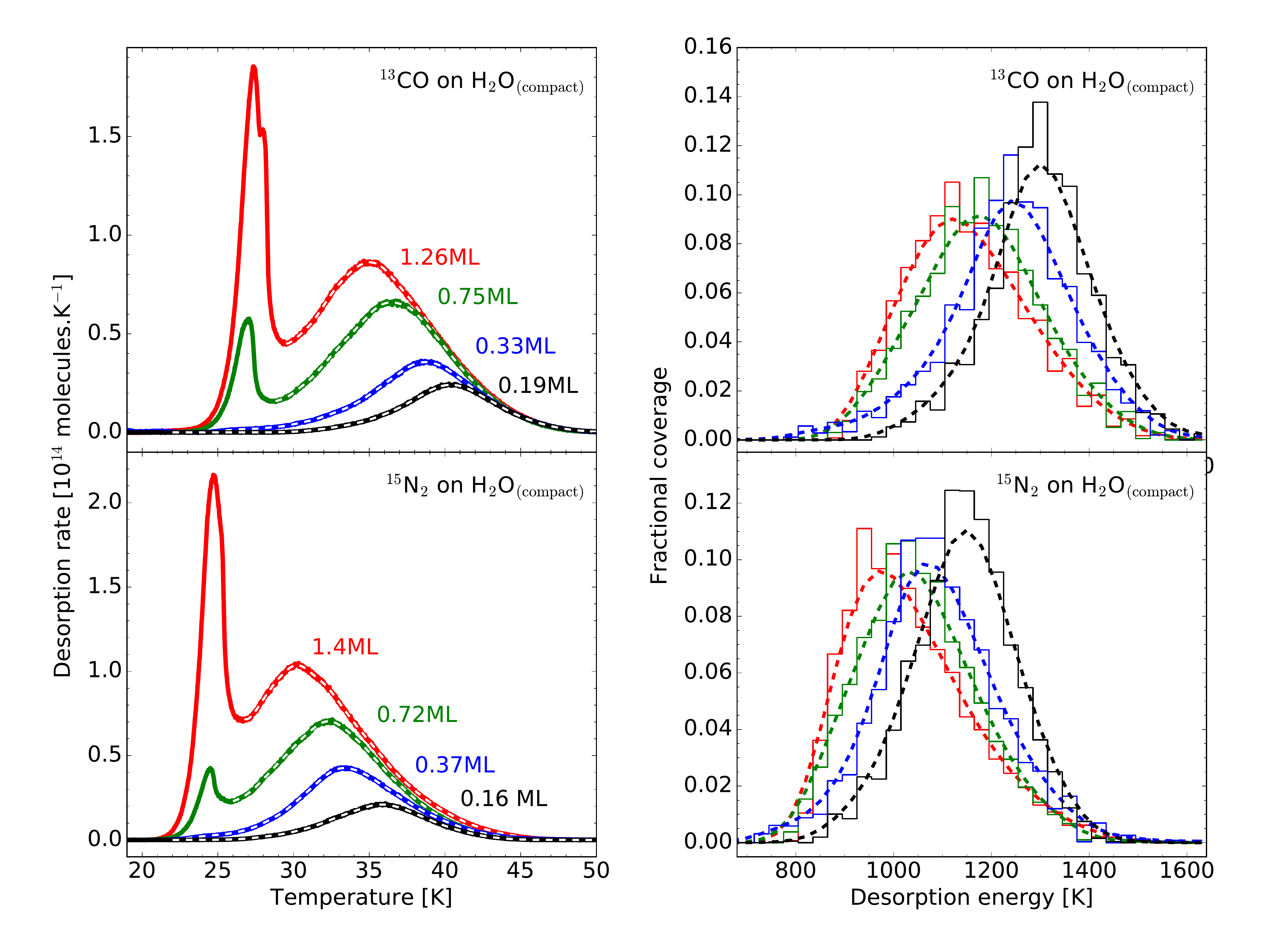}
      \caption{$^{13}$CO (left upper panel) and $^{15}$N$_2$ (left bottom panel) temperature desorption curves for various coverages on amorphous compact H$_2$O and the corresponding desorption energy distribution for $^{13}$CO (right upper panel) and $^{15}$N$_2$ (right bottom panel). On the left panel, the TPD data are the solid lines while the white dashed lines show the fit obtained with the corresponding energy distribution. In the right panel, the histograms show the fitted fractional coverage associated to a desorption energy and the dashed lines are the smoothed distribution using a gaussian filter for clarity. }\label{Fig_2}
\end{figure*}

}

\def\placefigurethree{

\begin{figure}[t]
  \centering
  \includegraphics[width=0.47\textwidth]{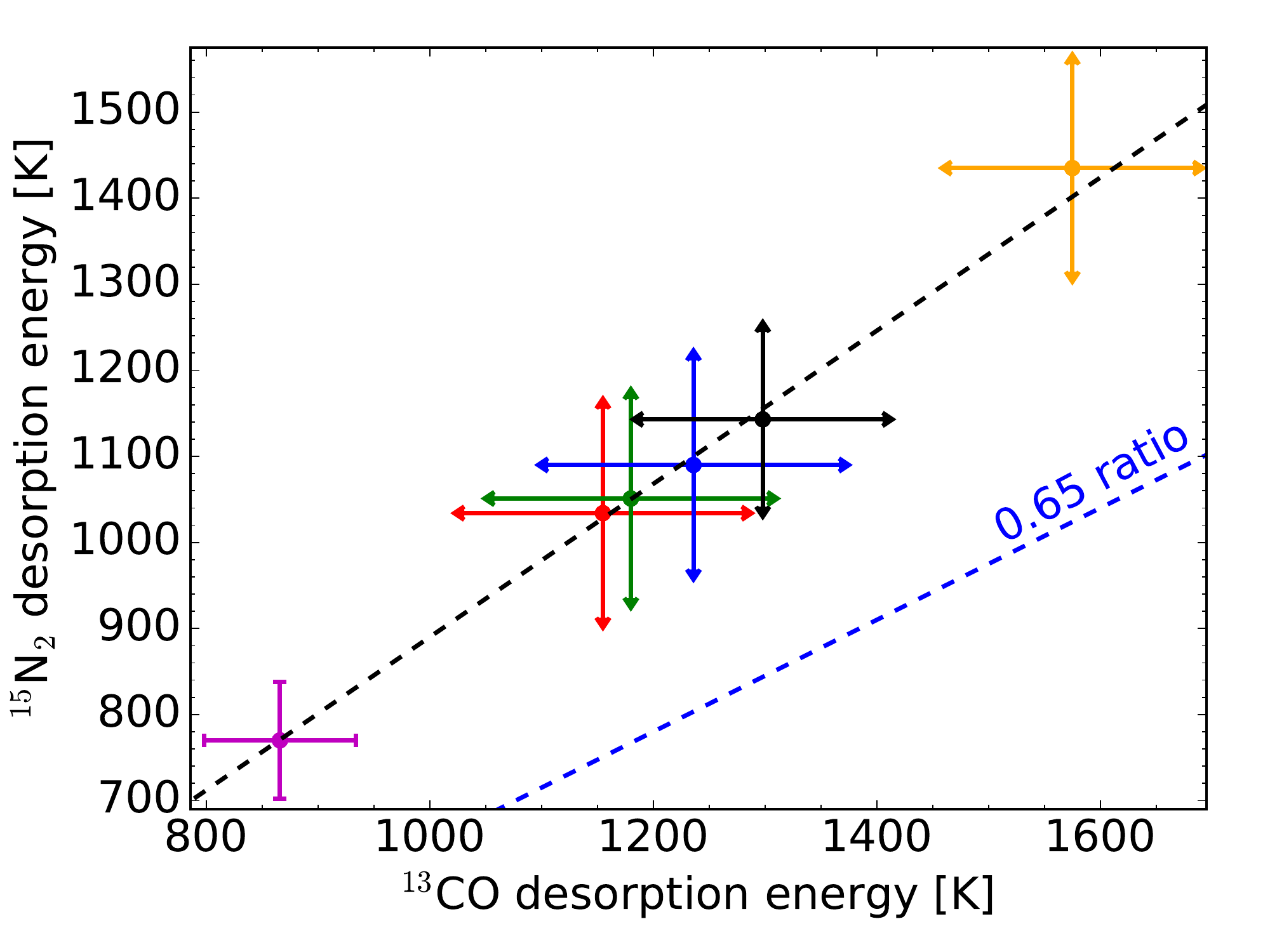}
    \caption{$^{15}$N$_2$ versus $^{13}$CO desorption energies and associated error bar for the pure ice multilayer regime (magenta symbols) and mean desorption energies for the submonolayer regimes on water substrates (red, green, blue, and black for the 1.4-1.3~ML, 0.7-0.8~ML, 0.4-0.3~ML, and 0.2~ML coverages on amorphous compact water, and orange symbol for $^{15}$N$_2$ and $^{13}$CO deposited on amorphous porous water). The black dashed line represents the fitted ratio of 0.89 and the blue dashed line shows the 0.65 ratio assumed in some astrochemical models to explain the observed CO and N$_2$H$^+$ abundances \citep{1997ApJ...486..316B}.}\label{Fig_3}
\end{figure}

}




  
\begin{abstract}
The relative desorption energies of CO and N$_2$ are key to interpretations of observed interstellar CO and N$_2$ abundance patterns, including the well-documented CO and N$_2$H$^+$ anti-correlations in disks, protostars and molecular cloud cores. Based on laboratory experiments on pure CO and N$_2$ ice desorption, the difference between CO and N$_2$ desorption energies is small; the N$_2$-to-CO desorption energy ratio is 0.93$\pm$0.03. Interstellar ices are not pure, however, and in this study we explore the effect of water ice on the desorption energy ratio of the two molecules. We present temperature programmed desorption experiments of different coverages of $^{13}$CO and $^{15}$N$_2$ on porous and compact amorphous water ices and, for reference, of pure ices. In all experiments, $^{15}$N$_2$ desorption begins a few degrees before the onset of $^{13}$CO desorption. The $^{15}$N$_2$ and $^{13}$CO energy barriers are 770 and 866~K for the pure ices, 1034 --1143~K and 1155--1298~K for different sub-monolayer coverages on compact water ice, and 1435 and 1575~K for $\sim$1~ML of ice on top of porous water ice. For all equivalent experiments, the N$_2$-to-CO desorption energy ratio is consistently 0.9. Whenever CO and N$_2$ ice reside in similar ice environments (e.g. experience a similar degree of interaction with water ice) their desorption temperatures should thus be within a few degrees of one another. A smaller N$_2$-to-CO desorption energy ratio may be present in interstellar and circumstellar environments if the average CO ice molecules interacts more with water ice compared to the average N$_2$ molecules.

\end{abstract}

\keywords{astrochemistry -- ISM: abundances -- ISM: molecules -- molecular data -- molecular processes}


\section{Introduction}


The chemical structures of interstellar clouds, cloud cores, protostellar envelopes, and protoplanetary disks are all regulated by the differential freeze-out and desorption of the main carriers of oxygen, carbon and nitrogen \citep{1997ApJ...486..316B,2002AA...386..622A,2013ChRv..113.9016H}. The sequential freeze-out of atoms and molecules onto interstellar grains is the starting point for a rich surface chemistry that is, e.g., responsible for most of the water in space \citep{2013ChRv..113.9043V}, as well as the abundant existence of complex, saturated molecules such as HCOOCH$_3$ \citep{2008ApJ...682..283G}. Freeze-out also affects gas-phase compositions in multiple ways. For example, CO freeze-out is a prerequisite for abundant N$_2$H$^+$ in molecular clouds, protostars, and protoplanetary disks \citep{2002ApJ...570L.101B,2005AA...435..177J,2013ApJ...765...34Q}.

The balance of freeze-out and desorption in disks also affects several aspects of planet formation. 
Condensation fronts in the mid planes of protoplanetary disks, so called snowlines, can enhance the planet formation efficiency due to increased grain surface density, rapid particle growth due to cold-head effects, pressure traps and increased grain stickiness \citep{Ciesla:2006dm,2007Natur.448.1022J,2011Icar..214..717G,2013AA...552A.137R}. Snowline locations also regulate the compositions of forming planets \citep{2011ApJ...743L..16O} and planetesimals. The locations of major snowlines depend on the volatile composition (e.g. whether most nitrogen is in N$_2$ or NH$_3$), a balance between freeze-out and thermal and non-thermal desorption rates at different disk locations, and disk dynamics \citep{2011ApJ...734...98O,2012ApJ...747..138O,2014ApJ...793....9A,2015AA...577A..65B}. Two of the most important volatiles in disks (as well as in clouds and protostars, are CO and N$_2$. Their desorption kinetics, fundamentally set by their binding energies, will determine  the locations of two of the most important disk snowlines.

CO and N$_2$ binding energies have been the subject of several previous studies. In two related studies, \citet{Oberg:2005iw} and \citet{Bisschop:2006dc} found that the binding energies of CO and N$_2$ in pure, layered and mixed CO:N$_2$ ices were relatively similar, i.e., the ratio of the N$_2$ to CO binding energies were 0.93 -- 1. These experiments did not consider the effects of water. Experiments on CO deposited on water ice has shown that CO is substantially more strongly bound in water-dominated ices compared to pure CO ices \citep{2012MNRAS.421..768N}; \citet{Collings:2003jt} found a 40\% higher desorption energy for CO on top of low-density amorphous water ice compare to pure CO ice. There are no similarly detailed studies of N$_2$ interactions with water ice, but cluster calculations suggest that N$_2$ may not bind very strongly to water ice \citep{Sadlej:1995bw}. Based on those calculations N$_2$ desorption energies of $0.65\times E_{\rm des}(CO)$ are sometimes used in astrochemical studies \citep[e.g.][]{1997ApJ...486..316B}. Such a low N$_2$ binding energy compared to CO naturally explains the presence of N$_2$H$^+$ in cores and disks where CO has frozen out \citep[e.g.][]{2013ApJ...765...34Q}, but seems inconsistent with the experimentally measured small difference in binding energies of CO and N$_2$ in pure ices \citep{Bisschop:2006dc}. 

In this study we explore the effect of water on CO and N$_2$ desorption energies to astrophysically relevant ices. We aim to answer 1. whether the ratio of N$_2$ to CO binding energies in water-dominated ices deviate from the ratio of 0.93 found for pure ices, and 2. whether the relative binding energies of CO and N$_2$ in water-dominated ices depend on the exact ice environment. In \S2 we present the experimental method -- temperature programmed desorption -- used to characterize CO and N$_2$ desorption. The experimental results and the derived CO and N$_2$ binding energies are presented in \S3. The experimental results and their astrophysical implications are then discussed in \S4.
 
 \section{Methods}
\label{methods}

Temperature Programmed Desorption (TPD) experiments are used to derive the desorption energies of $^{13}$CO or $^{15}$N$_2$ ices on CsI and H$_2$O substrates. Ices are grown by injecting molecules through a 4.8~mm diameter pipe at 0.7~inch from the substrate on a CsI window, resulting in a uniform ice. The window can be cooled to $\sim$11~K using a close-cycle He cryostat, and is placed in an ultra-high vacuum chamber with base pressures of $<$5.10$^{-10}$~Torr at room temperature. More details on the experimental setup are given in \citet{2015ApJ...801..118L}. The vapor pressure of deionized water purified through at least three freeze--pump--thaw cycles using liquid nitrogen is deposited on the CsI window at i) $\sim$~100K to grow amorphous compact water ice substrates, and ii) 11K to grow amorphous porous water ice substrates. $^{13}$CO (99\% purity, Sigma-aldrich) and $^{15}$N$_2$ (98\% purity, Sigma-aldrich) gases are then deposited at 11~K on top of the chosen substrate. The amount of molecules deposited is monitored during the injection using a calibrated quadrupole mass spectrometer (Pfeiffer QMG 220M1), integrating the mass spectrometer signal over time. The ice coverage is given in monolayer units with the typical approximation of 1~ML = 10$^{15}$~molecules~cm$^{-2}$. The chamber is also equipped with a Fourier transform infrared spectrometer (Bruker Vertex 70v) in transmission mode to monitor the amount of infrared active molecules deposited on the window in the mid-infrared. $^{13}$CO or $^{15}$N$_2$ ices of the desired thickness are then heated at a constant rate of 1~K~min$^{-1}$.

The temperature controller used to monitor the temperature is coupled to a thermocouple attached on a metallic window holder (Lakeshore 335). It has a relative uncertainty of 0.1~K but the absolute temperature is more difficult to assess since it depends on the thermal contact with the window holder it is attached to. We calibrated the temperature against initial CO TPD data obtained by the setup when the thermal contact was excellent \citep{Cleeves:2014dv}, and for which the resulting CO desorption energy was within the average energy obtained in the literature \citep{Collings:2003jt,Bisschop:2006dc,Acharyya:2007gj,2010AA...522A.108M,2014AA...564A...8M,Collings:2015fs}. We estimate that there is a 2~K absolute uncertainty on the temperature, based on the spread in the CO desorption energies found in the literature. The desorbing molecules are monitored using a quadrupole mass spectrometer (Hiden IDP 300, Model HAL 301 S/3) equipped with a pinhole on a translation stage that is approached 0.5 inches away from the CsI window. $^{13}$CO and $^{15}$N$_2$ isotopologues ({\it m/z}=29 and {\it m/z}=30 respectively) are used to rule out possible contamination in the TPD results, due to background deposition of $^{12}$CO and $^{14}$N$_2$ ({\it m/z}=28 for both). Analysis of the TPD experiments showed that this contamination is minimal (lower than the purity percentage given by the manufacturer). The TPD plots in desorbing molecules per K are obtained by subtracting the mass background for $^{13}$CO or $^{15}$N$_2$ and scaling the QMS signal so the TPD integral over the temperature range is equal to the amount of molecules deposited. This assumes that the signal detected by the QMS is proportional to the amount of molecules desorbing and that the pumping speed in the chamber is high, both of which have been verified. \

\placefigureone

The experimental data set consists of various $^{13}$CO or $^{15}$N$_2$ coverages deposited on the CsI window, on $\sim$50~ML$_{eq}$ of compact amorphous water, and on $\sim$50~ML$_{eq}$ of porous amorphous water. The TPD curves are fit using the Polanyi-Wigner equation:

\begin{equation}
-\frac{{d} \rm \theta}{\mathrm{d}\rm T} = \frac{\nu}{\beta} \, \theta^n \,  \mathrm{ e}^{-E_{\rm des}/T} 
\label{eq_poly}
\end{equation} 
, where $\theta$ is the ice coverage, T the temperature in K, $\nu$ a pre-exponential factor in s$^{-1}$, $\beta$ the heating rate in K~s$^{-1}$, n the desorption order, and E$_{des}$ the desorption energy in K. To derive the desorption energies, we describe the desorption kinetics using two different regimes: a multilayer regime regulated by $^{13}$CO-$^{13}$CO or $^{15}$N$_2$-$^{15}$N$_2$ binding energies, resulting in a zeroth order kinetics ($n\, =\, 0$ in equation \ref{eq_poly}) and a sub-monolayer regime where $^{13}$CO or $^{15}$N$_2$ are in contact with the substrate, resulting in a first order desorption. The zeroth-order regime is usually well fit by only one single desorption energy and the sub-monolayer regime needs to be described using a distribution of desorption energies. This is due to the different adsorption sites from a disordered and rough substrate, as reported recently by \citep{2012MNRAS.421..768N,Doronin:2015he,Collings:2015fs}, using models based on work by \citep{Tait:2005de,Koch:1997iv,Redhead:1962df}. 
For the pre-exponential factor associated to $^{13}$CO and $^{15}$N$_2$, we use the harmonic oscillator relation 
\citep[e.g.][]{Hasegawa:1992bx,Acharyya:2007gj,2012MNRAS.421..768N}:

\begin{equation}
\nu = \sqrt{\frac{2\,n_s\, E_{des}}{\pi^2 \, m}},
\label{eq_preexp}
\end{equation}
where $n_s$ is the number of adsorption sites ($\sim$10$^{19}$ sites.m$^{-2}$) and m is the mass of the molecule in kg. This approximation is valid in the case of small molecules like CO and N$_2$, but is not appropriate for large molecules, since it relies on internal and translational degrees of freedom being equivalent for the adsorbed and desorbing molecule \citep[e.g.][]{Muller:2003fh}. The value of the pre-exponential factor affects the derived desorption energy values, but not the ratio of CO and N$_2$ desorption energies, even when the factor is varied over many orders of magnitude.

\placefiguretwo 

\section{Results}
\label{res}

The TPD curves of $^{13}$CO or $^{15}$N$_2$ on the CsI substrate, compact, and porous amorphous water are shown in Fig. \ref{Fig_1}. In the top panel, $\sim$5~ML of $^{13}$CO or $^{15}$N$_2$ are deposited on the CsI window, then warmed up at 1~K min$^{-1}$. The desorption peaks at temperatures of 24.9~K for $^{15}$N$_2$ and 28.2~K for $^{13}$CO. The shape of the curve is similar to a 0th order desorption with an irregular desorption tail, indicative of different sub-monolayer binding site on the CsI window and perhaps also the window holder.
The middle panel of figure \ref{Fig_1} shows the TPD curves of $\sim$0.7~ML of $^{13}$CO or $^{15}$N$_2$ deposited on an amorphous compact thick water ice ($\sim$ 50~ML) and warmed up at 1~K min$^{-1}$. The curve has two peaks, which can be attributed to desorption 
from pure ice and desorption from the water substrate. The second peak, associated with the sub-monolayer interaction of the diatomic species with H$_2$O, is much broader than the desorption in the multilayer regime and peaks at 
32.4~K for $^{15}$N$_2$ and at 36.6~K for $^{13}$CO. The bottom panel shows the TPD curves of 2~ML of $^{13}$CO or $^{15}$N$_2$ deposited on amorphous porous water. From the porous H$_2$O ice (T$_{\rm deposition}$=11~K), $^{13}$CO and $^{15}$N$_2$ present two desorption peaks, one at 43 and 47~K, respectively, and one close to the water desorption temperature due to release of entrapped molecules (not shown here). This latter feature is due to volatile entrapment 
within the ice pores \citep[e.g.][]{Collings:2003kx,BARNUN:2007dc,2011AA...529A..74F,2014AA...564A...8M}.
0.7~ML - 0.8~ML of $^{13}$CO and $^{15}$N$_2$ desorbed below 65~K while the rest was entrapped with H$_2$O.
\placetableone

The experiments described above clearly demonstrate that $^{15}$N$_2$ desorption behavior is strongly affected by the presence of water, similarly to what has previously been observed for CO. The temperature shifts between pure ice desorption from compact and porous amorphous water ice appear to be similar for the two molecules.

To better comprehend the effect of the water substrate on $^{13}$CO and $^{15}$N$_2$ desorption kinetics, we performed a series of TPD experiments for different coverages on amorphous compact water. The results are shown in the left panels of figure \ref{Fig_2}. For high coverages above 0.7~ML, both a multilayer and a submonolayer component are present in the TPD curves. Note that the presence of a multilayer component below one monolayer could either be due to 
a lower surface site density 
than the assumed value of 10$^{15}$ molecules cm$^{-2}$ or 
to a non uniform wetting of the surface resulting in the formation of islands. 
For lower coverages, only the sub-monolayer desorption peak is present. For both $^{13}$CO and $^{15}$N$_2$, the sub-monolayer peak shifts towards higher temperatures with decreasing coverage. This coverage trend on compact amorphous water was also observed by \cite{2012MNRAS.421..768N} for CO, who explained it by CO adsorbing first to the most strongly bond sites on the water substrate.\

To quantify the desorption energy of these systems, we fit the TPD curves using the Polanyi-Wigner equation (eq. \ref{eq_poly} in section \ref{methods}), assuming a zeroth order for the 5~ML experiments on bare CsI window. We fit the logarithm of the desorbing molecules versus the inverse of the temperature with a straight line \citep[e.g.][]{Doronin:2015he}, yielding desorption energies for pure ices of 770 $\pm$ 68~K for $^{15}$N$_2$, and 866 $\pm$ 68~K for $^{13}$CO 
(Table \ref{Tab_1}, first row). The associated error mainly comes from the absolute uncertainty on the temperature while the relative uncertainty on the fit is less than 5~K. These desorption energy values result in pre-exponential factor values of 6.5~$\times$~10$^{11}$ s$^{-1}$ for $^{15}$N$_2$ and 7.1~$\times$~10$^{11}$ s$^{-1}$ for $^{13}$CO using equation \ref{eq_preexp}. Note that an empirical determination of the pre-exponential factor, using the intercept of the straight line fitting explained above yields values of 6.6~$\times$~10$^{11}$ s$^{-1}$ for $^{15}$N$_2$ and 8.0~$\times$~10$^{11}$ s$^{-1}$ for $^{13}$CO, which is in good agreement with the theoretical value. The desorption energies are consistent with literature data from \cite{Oberg:2005iw} of 790 $\pm$ 25~K and 855 $\pm$ 25~K for N$_2$ and CO, from \cite{2012MNRAS.421..768N} of 828 $\pm$ 28~K for CO, and CO desorption energies from \cite{Collings:2015fs} of 830 $\pm$ 36~K. Our N$_2$ desorption energy is substantially lower, however, than their measured value of 878 $\pm$ 36~K.

To derive desorption energies for the submonolayer regime, we used a distribution of binding energies obtained by fitting the sub-monolayer regime of the TPD curves by a linear combination of first order kinetics, sampling the desorption energy by steps of 30~K between 670~K and 1650~K. This technique takes into account the non-homogeneous nature of the amorphous water ice surface and has been recently used by \citet{Doronin:2015he} in the case of methanol adsorbed on graphite. The fitting is done in python using {\it scipy.optimize.nnls}, a non-negative least square fitting module, so the linear combination coefficients are kept positive. The linear combination coefficients are normalized to the initial coverages, yielding fractional coverages, and are plotted in the right panels of figure \ref{Fig_2} versus the sampled desorption energy. The data are smoothed using a gaussian filter and plotted in dashed lines for clarity as well. 
All the distributions are close to symmetric around the mean and present full width half maxima of 113 --139~K, resulting in well-defined 'representative' desorption energies for each coverage. The mean energy is known with a $\sim$30~K relative precision due to the chosen sampling energy steps (the uncertainty from the fit is smaller), and has an absolute uncertainty of 67~K.
As noted for the TPD curves, the desorption distribution and their trends for various coverages are similar for $^{13}$CO and $^{15}$N$_2$. 
The mean desorption energy values and the full width half maximum are reported in Table \ref{Tab_1}. The mean submonolayer desorption energy from a compact water ice surface ranges from 1034 to 1143~K for $^{15}$N$_2$ and 1155 to 1575~K for $^{13}$CO. The shift towards higher energy for decreasing submonolayer coverages is consistent with data from \cite{2012MNRAS.421..768N} for CO on amorphous compact water. The resulting pre-exponential factor values from equation \ref{eq_preexp} are between 6 and 10~$\times$~10$^{11}$ s$^{-1}$ over the sampled desorption energy range.

The desorption energies obtained for all the $^{15}$N$_2$ experiments are plotted versus those of $^{13}$CO in Fig. \ref{Fig_3}. The data is consistent with a single ratio of $\sim$0.9 and inconsistent with a ratio of 0.65. The desorption energy ratio of $^{15}$N$_2$ over $^{13}$CO for the multilayer (pure ice) and the mean desorption energy ratio of $^{15}$N$_2$ over $^{13}$CO for the sub-monolayer coverages 
are also listed in Table \ref{Tab_1}; the values span 0.88 -- 0.91. 


\placefigurethree

\section{Astrophysical implications}

The locations of condensation fronts (snowlines) in disks, protostars and clouds depend on the desorption energies of the volatiles in question. 
These desorption energies increase dramatically when CO or N$_2$ desorbs from water ice. The highest desorption energies barriers of  $\sim$1610 and 1470~K for CO and N$_2$, respectively, are achieved when CO and N$_2$ are deposited on a porous water ice surface where each CO or N$_2$ molecules can interact with multiple H$_2$O molecules. The large effects of water ice on the CO desorption energy was known from previous experiments \citep{Collings:2003kx,2012MNRAS.421..768N}. Our study shows that N$_2$ is similarly affected, and both molecules are therefore likely to present ranges of thermal desorption temperatures in different interstellar and circumstellar environments. In a typical protoplanetary disk a change in desorption energy from 770 to 1435~K results in a change in N$_2$ snowline location from $\sim$50 AU to $\sim$20~AU. This estimate is based on the median temperature disk profile $T=200{\rm K} \times (r/{\rm 1\;AU})^{-0.62}$ from \citet{2007ApJ...659..705A} and using the prescription from \citet{2009ApJ...690.1497H} to calculate the sublimation temperatures from the desorption energies. In the Solar Nebula this difference in N$_2$ snowline location between 50 and 20~AU is the difference between comets and the Ice Giants forming nitrogen rich or nitrogen poor \citep{2005Natur.435..459T}. 

The second astrophysical important result of our experiments is the similarity of the N$_2$ and CO desorption kinetics and 
energies in different ice environments. Whether the ices are pure or deposited on top of different kinds of amorphous water ices, and whether the ices are a more than a monolayer thick or a fraction of monolayer, the ratio between the N$_2$ and CO desorption energies is consistently 0.9. This implies that in astrophysical environments where CO and N$_2$ ices experience similar levels of interaction with water ice, the N$_2$ desorption energy and temperature can always be parameterized as a fraction (0.9) of the CO desorption energy and temperature.  

While the N$_2$-to-CO desorption energy ratio is certainly not unity, it is not close to the value of 0.65 preferred in some observational studies. In cloud cores different formation kinetics of CO and N$_2$ in the gas-phase may be sufficient to explain the later freeze-out of N$_2$, but in disks, where high densities result in short chemical times scale it is less clear that a N$_2$-to-CO desorption energy ratio of 0.9 is sufficient to explain observed N$_2$H$^+$ emission exterior to the CO snowline.



It is plausible, however, that N$_2$ on average is interacting with 
less H$_2$O-rich environment than CO. \citet{2011ApJ...735...15G} finds that the H$_2$O abundance in ices decreases with increasing ice coverage when modeling ice formation in dark clouds, i.e. the number of H$_2$O molecules in a specific ice layer is less in the upper layers of the ice mantle that formed at a later time compared to the lower layers of the ice mantle. There are two reasons why N$_2$ ice may form slightly later than CO ice and thus be mainly present in the top-most, water-poor ice layers. First the N$_2$ desorption temperature is slightly lower, which may be sufficient to keep N$_2$ in the gas-phase at lower temperature than CO if the freeze-out time-scales are long enough. Second, the nitrogen chemistry is slower compared to the CO one, which may 
cause N$_2$ ice to preferentially form later than CO ice 
\citep{2010AA...513A..41H,Pagani:2012da}. 
Both effects 
could contribute to the observed CO and N$_2$H$^+$ anti-correlation in molecular cloud cores. 
In disks, where the gas-phase chemistry is expected to reach steady state quickly, the different gas-phase time scales of N$_2$ and CO are not expected to play a role, but a slight differential freeze-out could. Differential freeze-out of CO and N$_2$ may also result in a high non-thermal desorption efficiency of N$_2$ compared to CO. \citet{2013ApJ...779..120B} found that N$_2$ ice UV photodesorption is very efficient and CO photodesorption is quenched if a CO ice is covered by a few 
N$_2$ ice layers. 


In summary, 
both CO and N$_2$ ice thermal desorption depend strongly on the ice morphology and composition.
Based on our experiments, N$_2$ and CO desorption energies are substantially elevated, when molecules are desorbing from an amorphous water ice surface compared to a pure ice. As long as this morphology and composition are equivalent for the two molecules, the N$_2$ desorption energy is 0.9 that of the CO desorption energy.  
Differential freeze-out may increase the difference, but detailed modeling is required to assess the feasibility of this scenario.

\acknowledgments
The authors are grateful to the referee for comments and suggestions that greatly improved the manuscript. E.C.F. is supported by a Rubicon fellowship (680-50-1302), awarded by the Netherlands Organisation for Scientific Research (NWO). K.I.\"O. acknowledges funding from the Alfred P. Sloan Foundation, and the Packard Foundation.
\bibliographystyle{apj}

\end{document}